\documentclass[12pt]{article}
\usepackage{fullpage,hyperref,bbm,amssymb,amsfonts,amsthm,amsmath}
\newcommand{\ket}[1]{|#1\rangle}
\newcommand{\bra}[1]{\langle #1|}
\newcommand{\<}{\langle}
\renewcommand{\>}{\rangle}

\newcommand{\F}{\mathbb{F}}
\newcommand{\Z}{\mathbb{F}}
\newcommand{\DFT}{\mathrm{DFT}}
\newcommand{\Tr}{\mathrm{Tr}}
\theoremstyle{plain}

\newtheorem{thm}{Theorem}[section]
\newtheorem{lm}[thm]{Lemma}
\newtheorem{cor}[thm]{Corollary}
\theoremstyle{remark}
\newtheorem{rem}[thm]{Remark}
\theoremstyle{definition}
\newtheorem{definition}[thm]{Definition}
\newtheorem{exa}[thm]{Example}
\newtheorem{prop}[thm]{Proposition}
\newcommand{\ZZ}{{\mathbb Z}}
\newcommand{\QQ}{{\mathbb Q}}
\newcommand{\NN}{{\mathbb N}}
\renewcommand{\AA}{{\mathbb A}}
\newcommand{\Sxw}{S^x_w}

\newcommand{\Xg}{X_\mathrm{good}}
\newcommand{\Wxg}{W^x_\mathrm{good}}
\newcommand{\Bxg}{B^x_\mathrm{good}}
\newcommand{\FF}{\mathbb{F}}
\newtheoremstyle{dotless}{}{}{\itshape}{}{\bfseries}{}{ }
		{\thmname{#1}\thmnumber{ #2}\thmnote{ (#3)}:}

\date{September 2, 2008}

\author{Thomas Decker\thanks{School of Computer Science, McGill
University, 3480 University Street, Montreal, Quebec H3A 2A7, Canada.
Electronic address: \texttt{decker@ira.uka.de}} \quad Jan
Draisma\thanks{Department of Mathematics and Computer Science,
Technische Universiteit Eindhoven, PO Box 513, 5600 MB Eindhoven, The
Netherlands. Electronic address: \texttt{j.draisma@tue.nl}} 
\thanks{Centrum voor Wiskunde en Informatica, Amsterdam, The Netherlands.}
\quad Pawel
Wocjan\thanks{School of Electrical Engineering and Computer Science,
University of Central Florida, Orlando, FL~32816, USA.  Electronic
address: \texttt{wocjan@cs.ucf.edu}}}

\begin{document}
\title{Efficient Quantum Algorithm for\\ Identifying Hidden Polynomials}

\maketitle

\abstract{We consider a natural generalization of an abelian Hidden
Subgroup Problem where the subgroups and their cosets correspond to
graphs of linear functions over a finite field $\F$ with $d$ elements.
The hidden functions of the generalized problem are not restricted to
be linear but can also be $m$-variate polynomial functions of total
degree $n\geq 2$.

The problem of identifying hidden $m$-variate polynomials of degree
less or equal to $n$ for fixed $n$ and $m$ is hard on a classical
computer since $\Omega(\sqrt{d})$ black-box queries are required to
guarantee a constant success probability.  In contrast, we present a
quantum algorithm that correctly identifies such hidden polynomials for
all but a finite number of values of $d$ with constant probability and
that has a running time that is only polylogarithmic in $d$.}

\section{Introduction}
Shor's algorithm for factoring integers and calculating discrete
logarithms \cite{Shor97} is one of the most important and well known
example of an exponential speed-up based on quantum computation. This
algorithm as well as other fast quantum algorithms for
number-theoretical problems \cite{Hallgren02,Hallgren05, SV05,
  Kedlaya} essentially rely on the efficient solution of an abelian
Hidden Subgroup Problem (HSP) \cite{BL95}.  This has naturally raised
the questions of what interesting problems can be reduced to the
non-abelian HSP and of whether the general non-abelian HSP can also be
solved efficiently on a quantum computer.

It is known that an efficient quantum algorithm for the dihedral HSP
would give rise to efficient quantum algorithms for certain lattice
problems \cite{Regev02}, and that an efficient quantum algorithm for
the symmetric group would give rise to an efficient quantum algorithm
for the graph isomorphism problem \cite{EH99}. Despite the fact that
efficient algorithms have been developed for several non-abelian HSP's
(see, for example, Ref.~\cite{ISS07} and the references therein), the
HSP over the dihedral group and the symmetric group have withstood all
attempts so far. Moreover, there is evidence that the non-abelian HSP
might be hard for some groups such as the symmetric group
\cite{Hallgren}.

Another idea to generalize abelian HSP is to consider
Hidden Shift Problems \cite{BD07,DHI:03} or problems with hidden
non-linear structures \cite{CSV07,HRS05,SR04}. In the latter context,
we define and analyze a black-box problem that is based on polynomial
functions of degree $n\ge 2$ and that can be reduced to an instance of
the yet unsolved Hidden Polynomial Problem (HPP)
\cite{CSV07}. Although our problem can be seen as a special case we
refer to it as HPP in the following.  The subgroups and the cosets of
the HSP are generalized to graphs of polynomial multivariate functions
going through the origin and to translated function graphs,
respectively.

To solve this new problem, we use the ``pretty good measurement''
framework, which was introduced in Ref.~\cite{BCD05} to obtain
efficient quantum algorithms for the HSP over some semidirect product
groups.  First, we reduce the HPP to a quantum state identification
problem. Second, we design a measurement scheme for distinguishing the
states. Third, we relate the success probability and implementation to
a classical algebro-geometric problem. The analysis of this classical
problem leads us to an efficient quantum algorithm for the black-box
problem.

This paper is organized as follows. In Section~2 we define the Hidden
Polynomial Problem and show that it suffices to solve the univariate
case on a quantum computer. In Section~3 we reduce this case to a
state distinguishing problem and present a measurement scheme to solve
it.  In Section~4, we prove that the measurement scheme can be
implemented efficiently and its success probability is bounded from
below by a constant, which is independent of $d$.  To do this, we
analyze the properties of an algebro-geometric problem related to the
black-box problem. In Section~5 we conclude and discuss possible
objectives for further research.

\section{Hidden Polynomial Problem}
The Hidden Polynomial Problem is a natural generalization of the
abelian HSP over groups of the special form $G:=\F^{m+1}$. The hidden
subgroup is defined by the $m$ generators $(0, \ldots, 1 , \ldots, 0
,q_i) \in \Z^{m+1}$ where the $1$ is in the $i$th component and $q_i$
is in $\Z$. In this case, the hidden subgroup $H_Q$ and its cosets
$H_{Q,z}$ for $z\in\Z$ are given by
\[
H_Q:=\{(x,Q(x)): x \in \Z^m \} \quad\mbox{and}\quad
H_{Q,z}:=\{(x,Q(x)+z): x \in \Z^m \}
\]
where $Q$ is the unknown linear polynomial $Q(X_1,\ldots,X_m)=q_1 X_1
+ \ldots + q_m X_m$. For the HPP we also consider polynomials of
higher degree.

\begin{definition}\label{DefProb}${}$
Let $\F$ be a finite field with $d$ elements and characteristic $p$
and let $Q(X_1, \ldots, X_m)\in \F[X_1, \ldots, X_m]$ be an arbitrary
polynomial with total degree $\deg(Q) \leq n$ and vanishing constant
term\footnote{A polynomial with constant term could also be considered
  in the following discussions.  However, the constant term is
  randomized by our algorithm and cannot be determined as a
  consequence.}.  Furthermore, let $B : \F^{m+1} \rightarrow \F$ be a
black-box function with
\[
B(r_1, \ldots, r_m ,s):=\pi(s-Q(r_1, \ldots, r_m))
\]
where $\pi$ is an unknown (but fixed) arbitrary permutation of the elements of
$\F$.  The Hidden Polynomial Problem is to identify the
polynomial $Q$ if only the black-box function $B$ is given.
\end{definition}

\begin{rem}[General Definition of HPP]
The general HPP, which is defined in Ref.~\cite{CSV07}, can be
equivalently reformulated as follows: The black-box function $h :
\Z^\ell \rightarrow \Z$ is given by $h(r_1, \ldots, r_\ell):=\pi(P(r_1,
\ldots, r_\ell))$, where $\sigma$ is an unknown (but fixed) arbitrary
permutation of $\Z$ and $P(X_1, \ldots, X_\ell)$ is the hidden
polynomial. Hence, the black-boxes $B$ from Def.~\ref{DefProb} occur
as special cases when the polynomials $P$ are restricted to have the form
\[
P(X_1, \ldots, X_m, Y):= Y - Q(X_1, \ldots, X_m)\,. 
\]
This restriction makes it possible to obtain an efficient quantum algorithm.
\end{rem}

\begin{rem}[Classical Query Complexity]
To derive a lower bound on the classical query complexity, we only
consider the case of univariate polynomials of degree $1$. Due to the
permutation $\pi$ the function values $B(r,s)$ themselves are useless.
We need to obtain at least one collision, i.e., two different points
$(r,s)$ and $(\tilde{r},\tilde{s})$ with
$B(r,s)=B(\tilde{r},\tilde{s})$, to determine the slope of the hidden
line.  Assume we have queried the black-box $B$ at $N$ different
points and have not seen any collision.  Then we can exclude at most
$\binom{N}{2}=O(N^2)$ different slopes. Since there are $d$ different
slopes and all are equally likely, we have to make
$\Omega({\sqrt{d}})$ queries to determine the slope with constant
success probability.
\end{rem} 

We say that a quantum algorithm for this problem is efficient if its
running time is polylogarithmic in the field size $d$ for a fixed
number $m$ of variables and a fixed maximum total degree $n$.  We
present such an efficient algorithm by first classically reducing the
$m$-variate problem to the univariate problem and then by solving the
univariate on a quantum computer.  The reduction is described in the
following lemma.  For simplicity, we initially assume that the
univariate case can be solved with probability 1 and show then how
to deal with the other cases. 

\begin{lm}
Assume that we can solve the univariate problem of degree $n$ or less
with success probability $1$.  Then, there is a simple recursive
interpolation scheme that solves the $m$-variate problem by 
solving of at most
\begin{equation}\label{eq:numberOfUnivariate}
\kappa_m = n^{m-1} + n^{m-2} + \ldots + 1 
\end{equation}
univariate problems.
\end{lm}
\begin{proof}
First, rewrite $Q$ as
\[
Q(X_1, \ldots, X_m)=\sum_\alpha Q_\alpha(X_m) \cdot
X_1^{\alpha_1}\cdot \ldots \cdot X_{m-1}^{\alpha_{m-1}}
\]
where $\alpha=(\alpha_1, \ldots, \alpha_{m-1})$ is a vector with the
exponents of the variables $X_1,\ldots,X_{m-1}$. For the
recursion we assume that we have an efficient algorithm for
polynomials with $m-1$ variables or less. Then we solve the
$m$-variate problem with the following two steps.
\begin{itemize}
\item Step 1: Set the variables $X_1, \ldots, X_{m-1}$ to $0$. We obtain
\[
Q(0,\ldots, 0,X_m)=Q_{(0,\ldots, 0)}(X_m)\,,
\]
which is a univariate polynomial. It has no constant term because $Q$
also has no constant term. This is a univariate problem and can be
solved by assumption.

\item Step 2: For $n$ different fixed $t_j \in \F$ we
consider\footnote{Note that the degree of each variable in the
polynomials is w.l.o.g.  smaller than the size $d$ of $\F$ after
reducing exponents modulo $d-1$, which is the order of the
multiplicative group $\F^\times$.} the polynomials
\[
Q(X_1, \ldots, X_{m-1},t_j)=\sum_\alpha Q_\alpha(t_j) \cdot
X_1^{\alpha_1}\cdot \ldots \cdot X_{m-1}^{\alpha_{m-1}}
\]
where $Q_\alpha(t_j)$ is a constant coefficient. By assumption we can
determine all $Q_\alpha(t_j)$ for $\alpha\not=(0,\ldots, 0)$.  Denote
by $|\alpha|=\sum_j \alpha_j$ the degree of the monomial defined by
$\alpha$. Since for $|\alpha|\geq 1$ the polynomial $Q_\alpha(X_m)$
has degree $n-|\alpha|$ and since we know $n$ function values, we can
determine $Q_\alpha$ efficiently with Lagrange
interpolation \cite{Gathen}.
\end{itemize}
Let $\kappa_m$ be the total number of univariate problems with degree
$n$ or less that we have to solve in the recursive scheme.  We
have $\kappa_1=1$ and $\kappa_m = \kappa_1 + n \cdot
\kappa_{m-1}$. This leads to the expression in
Eq.~(\ref{eq:numberOfUnivariate}).
\end{proof}

We have assumed that the univariate case can be solved with success
probability $1$.  However, our quantum algorithm fails to correctly identify
the hidden univariate polynomial with some nonzero (but constant) probability $p_f$.  We can reduce the failure probability of the
quantum algorithm for the univariate case to $p_f/\kappa_m$ by
repeating it a certain number of times, which is independent of
$d$. Then, by the union bound we see that the failure probability of
the overall algorithm for the $m$-variate problem is at most $p_f$.

\section{Distinguishing Polynomial Function States}
Most quantum algorithms for HSP's are based on the standard approach,
which reduces black-box problems to state distinguishing problems.  We
apply this approach to the Hidden Polynomial Problem as follows:

\begin{itemize}
\item Evaluate the black-box function on an equally weighted
superposition of all $(r,s) \in \Z^2$.  The resulting state is
\[
\frac{1}{d} \sum_{r,s \in \Z}\ket{r}\otimes\ket{s}\otimes
\ket{\pi( s-Q(r))}
\]
\item Measure and discard the third register.  Assume we have obtained
  the result $\pi(z)$ with $z:=s-Q(r)$.  Then the state on the first
  and second register is $\rho_{Q,z}:= \ket{\phi_{Q,z}}
  \bra{\phi_{Q,z}}$ where
\[ 
|\phi_{Q,z}\> := \frac{1}{\sqrt{d}} 
\sum_{r \in \Z} \ket{r} \otimes \ket{Q(r)+z}
\]
with the unknown polynomial $Q$, and $z$ is uniformly at random. The
corresponding density matrix is
\begin{equation}\label{EQ EQ}
\rho_Q := \frac{1}{d} \sum_{z\in\F} |\phi_{Q,z}\> \<\phi_{Q,z}|\,.
\end{equation}
\end{itemize}
We refer to the states $\rho_Q$ as {\em polynomial function
states}. We have to distinguish these states in order to solve the
black-box problem.

\subsection{Structure of Polynomial Function States}
To obtain a compact expressions for polynomial function 
states $\rho_Q$ we introduce the shift operator
\[
S_\Delta:=\sum_{x \in \F} \ket{\Delta + x}\bra{x}
\]
for $\Delta \in \F$, which directly leads to
\[
\rho_Q = \frac{1}{d^2} \sum_{b,c\in\F} |b\>\<c| \otimes
S_{Q(b)-Q(c)}\,.
\]
Now we use the fact that the shift operators $S_\Delta$ for all
$\Delta\in\F$ can be diagonalized simultaneously with the Fourier
transform
\[
\DFT_\F:= \frac{1}{\sqrt{d}} \sum_{x,y\in\F} \omega_p^{\Tr(x y)}
|x\>\<y|
\]
over $\F$, where $\Tr:\F \to \F_p$ is the trace map of the field
extension $\F/\F_p$ and $\omega_p:=e^{2\pi i/p}$ is a primitive
complex $p$th root of unity.  The Fourier transform $\DFT_\F$ can be
approximated to within error $\epsilon$ in time polynomial in
$\log(|\F|)$ and $\log(1/\epsilon)$ \cite{DHI:03}.  For
simplicity, we assume that it can be implemented perfectly (as the
error can be made exponentially small with polynomial resources only).
We have
\[
\DFT_\F \cdot S_\Delta \cdot \DFT_\F^\dagger = \sum_{x \in \F}
\omega_p^{\Tr(\Delta x)} |x\>\<x|\,.
\]
Consequently, the density matrices have the block diagonal form
\begin{eqnarray*}
\tilde{\rho}_Q & := & (I_d \otimes \DFT_\F) \cdot \rho_Q \cdot (I_d
\otimes \DFT_\F^\dagger) \\ &=& \frac{1}{d^2} \sum_{b,c,x\in\F} \chi
\Big( [Q(b)-Q(c)]x \Big) |b\>\<c| \otimes |x\>\<x|
\end{eqnarray*}
in the Fourier basis where we set $\chi(z):=\omega_p^{\Tr(z)}$ for all
$z\in\F$ and where $I_d$ denotes the identity matrix of size $d$.

By repeating the standard approach $k$ times for the same black-box
function $B$, we obtain the density matrix $\tilde{\rho}_Q^{\otimes
k}$. After rearranging the registers we can write
\begin{eqnarray*}
\tilde{\rho}_Q^{\otimes k} & = & \frac{1}{d^{2k}} \sum_{b,c,x\in\F^k}
\chi\!\left( \sum_{j=1}^k \left[Q(b_j)-Q(c_j)\right] x_j \right)\,
|b\>\<c| \otimes |x\>\<x| \\
& = & \frac{1}{d^{2k}} \sum_{b,c,x\in\F^k} \chi\!\left( \sum_{j=1}^k
\left[\sum_{i=1}^n q_i(b_j^i - c_j^i)\right]x_j \right)\, |b\>\<c|
\otimes |x\>\<x| \\
& = & \frac{1}{d^{2k}} \sum_{b,c,x\in\F^k} \chi\!\left( \sum_{i=1}^n
q_i \left[\sum_{j=1}^k (b_j^i - c_j^i)x_j\right] \right)\, |b\>\<c|
\otimes |x\>\<x| \\
& = & \frac{1}{d^{2k}} \sum_{b,c,x\in\F^k} 
\chi \Big( \big\< q, \big( \Phi_n(b) -\Phi_n(c) \big) x \big\> \Big) \, 
|b\>\<c| \otimes |x\>\<x|\,,
\end{eqnarray*}
where $q$, $\Phi_n(b)$, and $\Phi_n(c)$ are defined as follows:
\begin{itemize}
\item $q:=(q_1,q_2,\ldots,q_n)^T \in \F^n$ is the column
vector whose entries are the coefficients of the hidden polynomial
$Q(X)=\sum_{i=1}^n q_i X^i$
\item $\Phi_n(b)$ is the $n\times k$ matrix
\[
\Phi_n(b) := \sum_{i=1}^n \sum_{j=1}^k b_j^i |i\>\<j| = \left(
\begin{array}{cccc}
b_1    & b_2    & \cdots & b_k    \\
b_1^2  & b_2^2  & \cdots & b_k^2  \\
\vdots & \vdots &        & \vdots \\
b_1^n  & b_2^n  & \cdots & b_k^n  
\end{array}
\right)
\]
\item $\< \cdot, \cdot \>$ denotes the map $\F^n\times \F^n \rightarrow \F$ with $\< v, w \> = v_1 w_1 + \cdots v_n w_n$ for $v,w\in\F^n$.
\end{itemize}

\subsection{Algebro-Geometric Problem}\label{algebro}
We now show how to construct an orthogonal measurement for
distinguishing the states $\tilde{\rho}_Q^{\otimes k}$ by applying and
suitably modifying the ``pretty good measurement'' techniques
developed in \cite{BCD06,BCD05,BD07}.  Both the success probability
and the efficient implementation of our measurement are closely related to
the following algebro-geometric problem:  Consider the problem to
determine all $b\in\F^k$ for given $x\in \F^k $ and
$w\in\F^n$ such that $\Phi_n(b) \cdot x = w$, i.e.,
\begin{equation}\label{EQ LL}
\left(
\begin{array}{cccc}
b_1 & b_2 & \cdots & b_k \\ b_1^2 & b_2^2 & \cdots & b_k^2 \\ \vdots &
\vdots & & \vdots \\ b_1^n & b_2^n & \cdots & b_k^n
\end{array}
\right)\cdot \left(
\begin{array}{c}
x_1 \\ x_2 \\ \vdots \\ x_k
\end{array}
\right) = \left(
\begin{array}{c}
w_1 \\ w_2 \\ \vdots \\ w_n
\end{array}
\right)
\end{equation}
We denote the set of solutions to these polynomial equations and its
cardinality by
\[
S_w^x := \{ b\in \F^k \,:\, \Phi_n(b) \cdot x = w\} \quad\; {\rm and}
\quad\; \eta_w^x := |S_w^x|\,,
\]
respectively. We also define the quantum states $|S_w^x\>$ to be the
equally weighted superposition of all solutions
\[
|S_w^x\> := \frac{1}{\sqrt{\eta_w^x}} \sum_{b\in S_w^x} |b\>
\]
if $\eta_w^x > 0$ and $\ket{S_w^x}$ to be the zero vector otherwise.
Using this notation we can write the state $\tilde{\rho}_Q^{\otimes
k}$ as
\begin{equation}\label{eq:rhoQNewBasis}
\tilde{\rho}_Q^{\otimes k} = \frac{1}{d^{2k}}\sum_{x\in\F^k}
\sum_{w,v\in\F^n} \chi\Big(\<q,w\> - \<q,v\> \Big) \sqrt{\eta_w^x
\eta_v^x} \, |S_w^x\>\<S_v^x| \otimes |x\>\<x|\,.
\end{equation}

\subsection{Idealized Measurement for Identifying the States}\label{IDeal}
We first consider an idealized situation to explain the intuition
behind the measurement that we will use in the following sections to
solve the HPP efficiently. Assume that there is an efficient 
implementation of the unitary transformation $U_x$ that depends on $x$
and that satisfies the equation
\begin{equation}\label{eq:quantumSampling}
U_x |S_w^x\> = |w\>
\end{equation}
for all $(x,w)$ with $\eta_w^x>0$. Then, there is an efficient
measurement for identifying the polynomial states with success
probability
\begin{equation}\label{EQ:Erf}
\frac{1}{d^{2k+n}} \sum_{x\in\F^k} \left( \sum_{w \in \F^n}
\sqrt{\eta_w^x} \right)^2\,.
\end{equation}

For the proof, we observe that the block structure of the 
states $\tilde{\rho}_Q^{\otimes k}$ in
Eq.~(\ref{eq:rhoQNewBasis}) implies that we can measure the second
register in the computational basis without any loss of information.
The probability of obtaining a particular $x$ is
\[
\Tr\left(\tilde{\rho}_Q^{\otimes k} (I_{d^k}\otimes |x\>\<x|)\right) =
\frac{1}{d^{2k}} \sum_{w\in\F^n} \eta_w^x = \frac{1}{d^k}\,,
\]
i.e., we have the uniform distribution, and the resulting reduced
state is
\begin{equation}\label{eqeq3}
\tilde{\rho}_Q^x := \frac{1}{d^k} \sum_{w,v\in\F^n} \chi\Big(\<q,w\> -
\<q,v\>\Big) \sqrt{\eta_w^x \eta_v^x} |S_w^x\>\<S_v^x|\,.
\end{equation}
We now apply $U_x$ to the state ${\tilde \rho}_Q^x$ of Eq.~(\ref{eqeq3})
and obtain
\[
U_x \tilde{\rho}_Q^x U_x^\dagger = \frac{1}{d^k} \sum_{w,v\in\F^n}
\chi\Big(\<q,w\> - \<q,v\>\Big) \sqrt{\eta_w^x \eta_v^x} |w\>\<v|\,.
\]
After having applied the transform $U_x$, we measure in the Fourier
basis, i.e., we carry out the orthogonal measurement with respect to
the states
\begin{equation}\label{eq:fourierStates}
|\psi_{Q'}\> := \frac{1}{\sqrt{d^n}} \sum_{w \in \F^n}
\chi\Big(\<q',w\>\Big) |w\>\,
\end{equation}
where $q'$ ranges over all tuples in $\F^n$.  
Simple computations show that the probability for the correct
identification of the state ${\tilde \rho}_Q^x$ is
\begin{equation}\label{eq:probForSpecific_x}
\langle \psi_Q | {\tilde \rho}_Q^x | \psi_Q \rangle =
\frac{1}{d^{k+n}} \left( \sum_{w \in \F^n} \sqrt{\eta_w^x}
\right)^2\,.
\end{equation}
The probability of correctly identifying $Q$ is obtained by averaging,
i.e., summing the probabilities in Eq.~(\ref{eq:probForSpecific_x})
over all $x$ and multiplying the sum by $1/d^k$.  It is equal to the
the expression in Eq.~(\ref{EQ:Erf}). This completes the proof.

The problem with this idealized measurement is that there are pairs
$(x,w)$ where $\eta_w^x$ is in the order of $d$.  It is not clear how
to implement the unitary $U_x$ in Eq.~(\ref{eq:quantumSampling})
efficiently in these cases. In the next subsection we consider an
approximate version $V_x$ of $U_x$. This approximation guarantees that
$U_x|S^x_w\>=V_x|S^x_w\>$ is satisfied for pairs $(x,w)$ with $1\le
\eta^x_w \le D$ where $D$ is some constant. We show that $V_x$ can be
implemented efficiently and that the resulting approximate measurement
is good enough to identify the states with constant success
probability.

\subsection{Approximate Measurement}\label{Sec app}
In this and the following sections we set $k=n$, i.e., the number $k$
of copies equals the maximum degree $n$ of the hidden
polynomials. Furthermore, let $D$ be some positive integer that
depends on $n$ but not on $d$, let $\Xg \subseteq \F^n$ be some
subset, and for $x \in \Xg$ let $\Wxg$ be some subset of 
$\{ w \in \F^n\, | \, 1 \le \eta^x_w \le D \}$.
The number $D$ and the sets $\Xg$ and $\Wxg$ will be determined later. We define the
subset
\begin{eqnarray}
\Bxg & := & \{ b \in \F^n\, | \, \Phi_n(b) \cdot x = w \mbox{ for some $w\in\Wxg$}\}\label{EEeq2}
\end{eqnarray}
for all $x\in\Xg$. 

\begin{lm}\label{Aplemma}
Assume that there are efficient classical methods for testing
membership in $\Xg$ and $\Wxg$ and for enumerating the elements of
$\Sxw$ for given $x\in\Xg$ and $w\in\Wxg$.  Then there is an efficient
approximate measurement for identifying the states with success
probability bounded from below by
\begin{equation}\label{eq:approxLowerBound}
\frac{1}{d^{3n}} \cdot |\Xg| \cdot |W_{\rm good}|^2\,,
\end{equation}
where $|W_{\rm good}|:=\min_{x\in\Xg} |\Wxg|^2$.
\end{lm}

\begin{rem}\label{RemRem}
Note that the lower bound is a constant if $|\Xg|=\Omega(d^n)$ and
$|W_{\rm good}|=\Omega(d^n)$. We analyze the algebro-geometric problem
and show that all the above properties are satisfied and the
cardinalities of the sets are sufficiently large.
\end{rem}

\begin{proof}
Let us assume that we have obtained $x\in\Xg$ in the first measurement
step as described in Section~\ref{IDeal}. The probability of this event
is $|\Xg|/d^n$. We now discuss the approximate transformation $V_x$
and the resulting success probability.  Let $P_{\mathrm{good}}$ be the
projector onto the subspace spanned by $|b\>$ for all $b\in\Bxg$.
Clearly, the orthogonal measurement defined by $P_{\rm good}$ can be
carried out efficiently since membership in $\Wxg$ can be tested
efficiently.  The probability to be in the ``good'' subspace is
\[
\Tr\Big(P_{\mathrm{good}}\, \tilde{\rho}_Q^x P_{\mathrm{good}} \Big) =
\frac{|\Bxg|}{d^n}
\]
and the resulting reduced density operator is
\begin{equation}\label{eq:reducedRhoGood}
\tilde{\rho}_{Q,\mathrm{good}}^x := \frac{1}{|\Bxg|} \sum_{w,v\in\Wxg}
\chi\Big(\<q,w\> - \<q,v\>\Big) \sqrt{\eta_w^x \eta_v^x}
|S_w^x\>\<S_v^x|\,.
\end{equation}
In the following we use the fact that for $x\in\Xg$ and all $w\in\Wxg$
the cardinality $\eta_w^x$ is bounded from above by $D$ and 
that the elements of the sets
$S^x_w$ can be computed efficiently.  In this case we have an
efficiently computable bijection between $S_w^x$ and the set $\{ (w,j)
\,:\, j=0,\ldots,\eta_w^x-1 \}$.  This bijection is obtained by
sorting the elements of $S_w^x$ according to the lexicographic order
on $\F^n$ and associating to each $b\in S_w^x$ the unique
$j\in\{0,\ldots,\eta_w^x-1\}$ corresponding to its position in
$S_w^x$.

We now show how to implement the transformation $V_x$ efficiently,
which satisfies
\[
V_x |\Sxw\> = |w\>\,.
\]
\begin{itemize}
\item Implement a transformation with
\begin{equation}
|b\> \otimes |0\> \otimes |0\> \mapsto |w\> \otimes |j\> \otimes
|\eta^x_w\>
\end{equation}
for all $b\in\Bxg$.  To make it unitary we can simply map
all $b\not\in\Bxg$ onto some vectors that are orthogonal (e.g., by
simply flipping some additional qubit saying that they are bad).  Note
that $b$ and $x$ determine $j$ and $w$ uniquely and vice versa.
Furthermore, we can compute $w$ and $j$ efficiently since $\eta_w^x$
is bounded from above by $D$.  Consequently, this unitary acts
on the states $|\Sxw\>$ as follows
\begin{equation}
\frac{1}{\sqrt{\eta_w^x}} \sum_{b \in S_w^x} \ket{b} \otimes \ket{0}
\otimes \ket{0} \mapsto \frac{1}{\sqrt{\eta_w^x}} \ket{w} \otimes
\sum_{j=1}^{\eta^x_w} \ket{j} \otimes \ket{\eta_w^x}
\end{equation}

\item Apply the unitary 
\[
\sum_{\ell=0}^{\eta_w^x-1} (F_{\ell+1} \oplus I_{d^n-\ell-1}) \otimes
\ket{\ell}\bra{\ell} + \sum_{\ell=\eta_w^x}^{d^n-1} I_{d^n} \otimes
\ket{\ell}\bra{\ell}
\]
on the second and third register. This implements the embedded Fourier
transform $F_\ell$ of size $\ell$ controlled by the second register in
order to map the superposition of all $\ket{j}$ with $j \in \{0,
\ldots, \ell-1\}$ to $\ket{0}$. The resulting state is $\ket{w}
\otimes \ket{0} \otimes \ket{\eta_w^x}$.

\item Uncompute $\ket{\eta_w^x}$ in the third register with the help
of $w$ and $x$. This leads to the state $\ket{w}\otimes \ket{0}
\otimes \ket{0}$
\end{itemize}
We apply $V_x$ to the state of Eq.~(\ref{eq:reducedRhoGood}) and
obtain
\[
V_x \tilde{\rho}_{Q,\mathrm{good}}^x V_x^\dagger = \frac{1}{|\Bxg|}
\sum_{w,v\in\Wxg} \chi\Big(\<q,w\> - \<q,v\>\Big) \sqrt{\eta_w^x
\eta_v^x} |w\>\<v|\,.
\]
We now measure in the Fourier basis, i.e., we carry out the orthogonal
measurement with respect to the states $|\psi_{Q'}\>$ defined in
Eq.~(\ref{eq:fourierStates}).  Analogously to the ideal situation we
obtain that the probability for the correct detection of the state
${\tilde \rho}_Q^x$ is
\begin{equation}\label{eq:probForSpecific_x_Good}
\<\psi_Q | V_x \tilde{\rho}_{Q,\mathrm{good}}^x V_x^\dagger |\psi_Q \>
= \frac{1}{d^{n}} \frac{1}{|\Bxg|} \left( \sum_{w \in \Wxg}
\sqrt{\eta_w^x} \right)^2\,.
\end{equation}
The overall success probability is 
\begin{equation}
\frac{1}{d^n} \sum_{x\in\Xg} \frac{|\Bxg|}{d^n} \<\psi_Q | V_x
\tilde{\rho}_{Q,\mathrm{good}}^x V_x^\dagger |\psi_Q \> =
\frac{1}{d^{3n}} \sum_{x\in \Xg} \left( \sum_{w \in \Wxg}
\sqrt{\eta_w^x} \right)^2\,.
\end{equation}
The first factor $1/d^n$ is the probability that we obtain a specific
$x$.  The right most expression is clearly at least the expression in
Eq.~(\ref{eq:approxLowerBound}).
\end{proof}

\section{Analysis of the Algebro-Geometric Problem}
In this section we show that the cardinalities of the sets $\Xg$ and
$\Wxg$ in Lemma~\ref{Aplemma} are sufficiently large in the case $k=n$
for all $\F$ that satisfy certain constraints on the
characteristic. This guarantees that the success probability of the
approximate measurement in Section~\ref{Sec app} is bounded from below
by a constant that does not depend on the field size.

Although our classical algebro-geometric problem appears to be very
similar to the average-case problem in Ref.~\cite{BCD05} for the HSP
over semidirect product groups, the elementary arguments of Lemma~5
in Ref.~\cite{BCD05} cannot be applied in a straightforward way to
prove that the cardinalities of the sets $\Xg$ and $\Wxg$ in
Lemma~\ref{Aplemma} are sufficiently large. More precisely, in the
case of the HPP we obtain the first two moments 
\begin{eqnarray}
\mathbb{E}\left[\eta_w^x\right] & = & d^{k-n} \;\;{\rm and} \label{eq:firstMoment}
\\ \mathbb{E}\left[(\eta_w^x)^2\right] & = &
\mathbb{E}\left[\eta_w^x\right] + \frac{1}{d^{k+n}} \sum_{b \neq c}
\sum_{x \in \F^k} \delta\Big[ \big(\Phi_n(b)-\Phi_n(c)\big) x =
  (0,0,\ldots,0)^T \Big]
\end{eqnarray}
for the $\eta_w^x$.  Since we have $b\neq c$, there is an index $j'$
with $b_{j'}\neq c_{j'}$.  It is clear that for all $b_j$, $c_j$, and
$x_j$ with $j\neq j'$ we have at most one $x_{j'}$ such that the
condition in the square bracket is satisfied but it is not obvious
when this $x_{j'}$ exists. In contrast to the situation in
Ref.~\cite{BCD05}, this argument only
leads to a weak upper bound
\begin{equation}
\mathbb{E}\left[(\eta_w^x)^2\right] \le
\mathbb{E}\left[\eta_w^x\right] + \frac{1}{d^{n}} (d^k-1)
d^{k-1}
\end{equation}
on the second moment. Eq.~(\ref{eq:firstMoment}) implies that the number of copies should be at least $n$.  In this case, however, the upper bound on the second moment is $\Omega(d^{n-1})$.  Therefore, we cannot use the probabilistic arguments of Ref.~\cite{BCD05} to prove
that $\Xg$ and $\Wxg$ have the desired properties.

In the following, we choose an approach that does not rely on any
probabilistic arguments.  We present two different proofs based on
algebro-geometric techniques that also show that the approximative
measurement can be implemented efficiently. Both proofs differ
slightly in their scope: The first analysis applies if the
characteristic of $\F$ is larger than $k=n$ and the second if a
certain polynomial with integer coefficients does not vanish when
considered modulo the characteristic. Hence, the second analysis can
be used in some cases when the first analysis cannot be applied and
vice versa.

The notions and results of algebra and algebraic geometry that are
used in the proofs can be found in Ref.~\cite{Lang} as well as in
Refs.~\cite{LittleCoxOShea,Gathen,Harris}.

\subsection{First Analysis} \label{ssec:FirstAnalysis}
For the analysis of the implementation of $V_x$ and the success
probability of our algorithm for $k=n$ we define the $n$ polynomials
$f_j\in\F[X_1,\ldots,X_n,B_1,\ldots,B_n]$ as
\[
\left( \begin{array}{c}f_1 \\f_2 \\ \vdots \\ f_n \end{array}\right)
:= \left(\begin{array}{cccc} B_1 & B_2 & \cdots & B_n \\ B_1^2 & B_2^2
& \cdots & B_n^2 \\ \vdots & \vdots & & \vdots \\ B_1^n & B_2^n &
\cdots & B_n^n\end{array} \right)\cdot \left(\begin{array}{c} X_1 \\
X_2 \\ \vdots \\ X_n\end{array} \right)\,,
\]
where the product of the matrix and the vector corresponds to the
left-hand side of Eq.~(\ref{EQ LL}). Furthermore, let $f$ be the
$n$-tuple $f:=(f_1,\ldots,f_n)$, which defines a map from
$\F^n\times\F^n$ to $\F^n$ with $f(x,b)=(f_1(x,b),\ldots,f_n(x,b))$.
Using this notation, $\Sxw$ can be expressed as
\[
\Sxw=\{b\in\F^n \,:\, f(x,b)=w\}\quad {\rm with} \quad w \in \F^n\,.
\]
For a fixed $x$ the tuple $f$ defines a map from $\F^n$ to $\F^n$ and
the sets $\Sxw$ are the preimages of $w\in\F^n$ under this map.

Let $\overline{\F}$ denote the algebraic closure of $\F$. We also view
$f$ as a map from $\overline{\F}^n$ to $\overline{\F}^n$.  For given
$x,w\in\overline{\F}^n$, we refer to the subvariety
$\{b\in\overline{\F}^n\, | \, f(x,b)=w\}$ of $\overline{\F}^n$ as the
fiber of $f(x,\cdot)$ over $w$.  In the proposition below, we choose
the sets $\Xg$ and $\Wxg$ such that the fibers of $f(x,\cdot)$ over
$w$ are zero-dimensional.  This implies that the numbers $\eta^x_w$
are bounded from above by some constant $D$ for all $x\in\Xg$ and
$w\in\Wxg$ since the sets $S^x_w$ are equal to the intersections of
the fibers with $\F^n$.

\begin{prop}\label{lmlm}
Assume that the characteristic $p$ of $\F$ is strictly larger than
$n$, let $\Xg := (\F^\times)^n$, and for $x \in \Xg$ set 
\[ \Wxg:=\{w \in \FF^n\, | \, \text{ the fiber of
$f(x,\cdot)$ over $w$ is zero-dimensional and } \eta^x_w \geq 1\}. \]
Then the requirements of Lemma~\ref{Aplemma} are satisfied and we have
$|\Xg|=\Omega(d^n)$ and $|\Wxg|=\Omega(d^n)$.
\end{prop}

\begin{proof}
We find the solutions of the system $f(x,b)=w$ efficiently as follows:
We precompute generic reduced Gr\"obner bases with Buchberger's
algorithm for the lexicographic order \cite{Gathen,LittleCoxOShea},
i.e., we treat the coefficients of the polynomials in the variables
$b_i$ as rational expressions in the variables $x_i$ and $w_i$.
Whenever Buchberger's algorithm requires division by a rational
expression $E$ in the $x_i$ and $w_i$, we distinguish between the case
where $E$ remains nonzero upon specializing $x$ and $w$ and the case
where $E$ becomes zero upon specialization. This precomputation yields
a finite decision tree whose leaves correspond to all possible reduced
Gr\"obner bases. In each leaf we can decide whether the solution
variety of the system $f(x,b)=w$ is zero-dimensional, and if so we can
compute an upper bound on its cardinality. Choose $D$ to be the
maximum over all these upper bounds.

On input $(\F, x, w)$ we now find the corresponding Gr\"obner basis by
evaluating a bounded number of rational expressions that also only
needs a bounded number of field operations. From the Gr\"obner basis
we can read off whether the set of solutions, i.e., the fiber of
$f(x,\cdot)$ over $w$ is zero-dimensional.  If this is the case, the
set $S^x_w$ of all solutions $b \in \F^n$ can be computed by
iteratively solving a bounded number of univariate equations, which
again can be done efficiently. By construction, this set has
cardinality at most $D$.

We now show that $|\Wxg|=\Omega(d^n)$ for all $x\in\Xg$. Fix $x \in
\Xg$. On the open set $\hat{U}$ in $\overline{\FF}^n$ where all
coordinates $b_i$ are distinct, the differential $d \varphi$ of the
map $\varphi : \hat{U} \rightarrow \overline{\FF}^n$ sending $b$ to
$f(x,b)$ has full rank everywhere.  Indeed, at $b$ the differential of
this map sends $c \in \overline{\FF}^n$ to
\[ 
\left( \begin{array}{ccccc} 1 & & & & \\ & 2 & & & \\ & & 3 & & \\ 
	& & & \ddots & \\ & & & & n \end{array} \right) 
\left( \begin{array}{ccc}
1 & \ldots & 1\\
b_1 & \ldots & b_n\\
b_1^{2} & \ldots & b_n^2\\
\vdots & & \vdots\\
b_1^{n-1} & \ldots & b_n^{n-1}
\end{array} \right) 
\left( \begin{array}{ccc} c_1 & & \\ & \ddots & \\ & & c_n \end{array} \right)
\left( \begin{array}{c} x_1 \\ \vdots \\ x_n \end{array} \right). 
\]
Now the first matrix is invertible because the characteristic of $\FF$
is larger than $n$, and the second matrix is invertible because the
$b_i$ are distinct. Hence if $d|_b\, \varphi$ maps $c$ to $0$ then all
$c_i x_i$ are zero, and as $x \in (\FF^\times)^n$ we find $c=0$, i.e.,
$d|_b \varphi$ is injective.

This implies that the fibers of $\varphi$ over $w$ are all
zero-dimensional.\footnote{This is an elementary statement from
  algebraic geometry: If some fiber has positive dimension, then it
  contains a point $b$ where the tangent space to the fiber has
  positive dimension. This tangent space is then mapped to zero by
  $d|_b\varphi$, a contradiction to the injectivity of this linear
  map. For a concise introduction to the interplay between dimension
  and tangent spaces we refer to \cite[chapter 9, paragraph
    6]{LittleCoxOShea}.} Their cardinalities are bounded from above by
$D$.  Let $U$ denote the intersection of $\hat{U}$ with $\F^n$.  The
upper bound implies that the size of the image $\varphi(U)$ is at
least $|\varphi(U)|\ge |U|/D=\Omega(d^n)$.  Clearly, the fibers of
$f(x,\cdot)$ over $w$ are zero-dimensional for all $w\in\varphi(U)$
that do not lie in the image of the complement of $\hat{U}$ under the
map $f(x,\cdot)$.  This image is certainly contained in some
subvariety $\hat{I}_x \subseteq \overline{\F}^n$ defined over $\F$ of
dimension $n-1$ since ${\rm
  dim}(\overline{\F}^n\setminus\hat{U})=n-1$.  Hence, we can apply
Schwartz-Zippel's theorem (Prop.~98 in Ref.~\cite{Zippel}) and
conclude that the cardinality of the intersection $I_x$ of $\hat{I}_x$
with $\F^n$ is at most $\kappa d^{n-1}$. Here $\kappa$ is a uniform
upper bound on the degree of the equation defining $I_x$, which can
again be found by a generic Gr\"obner basis computation without
specifying $x$.  This completes the proof that for each $x\in\Xg$ the
number of $w$ such that the fiber of $f(x,\cdot)$ over $w$ is
zero-dimensional is $\Omega(d^n)$.

\end{proof}

With Lemma~\ref{Aplemma} the following corollary is a direct
consequence of Prop.~\ref{lmlm}.

\begin{cor}
For $p>n$ the approximative measurement
of Sec.~\ref{Sec app} can be implemented efficiently. Furthermore, 
for the success probability we have 
\begin{eqnarray*}
\frac{1}{d^{3n}} \sum_{x \in \FF^n} 
	\left(\sum_{w \in \FF^n} \sqrt{\eta^x_w}\right)^2 
& \ge &
\frac{1}{d^{3n}} \sum_{x \in (\FF^\times)^n} 
	\left(\sum_{w \in \varphi(U)\setminus I_x} \sqrt{\eta^x_w}\right)^2 \\
& \ge &
\frac{1}{d^{3n}} (d-1)^n \left(\frac{d (d-1) 
\cdots (d-n+1)}{D} - \kappa d^{n-1}\right)^2\\
&=& 1/D^2 - O(1/d)\,,
\end{eqnarray*}
which leads to a lower bound that does not depend on the field size $d$. 
\end{cor}

\subsection{Second Analysis}
The following general proposition allows us to make statements about
the size of the preimages of a general morphism $f:\AA^m \times \AA^n
\to \AA^n$ over an affine space $\AA$ independently of the underlying 
field $\F$. This morphism should be thought of as a family of
morphisms from the $n$-dimensional space $\AA^n$ to itself,
parameterized by $\AA^m$.

\begin{prop} \label{prop:FiniteDominant}
Consider a morphism $f:\AA^m \times \AA^n \rightarrow \AA^n$ over
$\ZZ$, that is, $f$ is given by an $n$-tuple $f=(f_1,\ldots,f_n)$ of
polynomials in $\ZZ[X,B]$, where $X=(X_1,\ldots,X_m)$ and
$B=(B_1,\ldots,B_n)$ are the coordinates on $\AA^m$ and on the first
copy of $\AA^n$, respectively. Suppose that the Jacobian determinant
$\det (\partial f_i/\partial B_j)_{ij}$ is a non-zero
element\footnote{This condition on $f$ says that generic morphisms in
  this family are dominant. When we work over algebraically closed
  fields $\FF$ this means that the image is dense in $\FF^n$. The
  proposition states that over finite fields the generic morphism
  still hits a large subset of $\FF^n$.}  of $\ZZ[X,B]$. Then there
exists a real number $\gamma$ with $0<\gamma\le 1$ and a non-zero
polynomial $g \in \ZZ[X]$ such that for all finite fields $\FF$ and
all $x \in \FF^m$ with $g(x) \neq 0$ when $g$ is considered as a
polynomial over $\F$ we have $|f(\{x\} \times \FF^n)| \geq \gamma
|\FF|^n$.
\end{prop}

\begin{proof}
By the condition on the Jacobian determinant $f_1,\ldots,f_n \in
\QQ(X,B)$ are algebraically independent over $\QQ(X)$.\footnote{If $P
  \in \QQ(X)[W_1,\ldots,W_n]$ is of minimal degree with
  $P(f)=P(f_1,\ldots,f_n)=0$, then differentiation with respect to
  $B_j$ and the chain rules gives $\sum_i \frac{\partial P}{\partial
    W_i}(f) \frac{\partial f_i}{\partial B_j}=0$, so that
  $(\frac{\partial P}{\partial W_i}(f))_i$ is in the row kernel of the
  Jacobian matrix, and non-zero by minimality of $\deg(P)$---whence
  $\det(\frac{\partial f_i}{\partial B_j})=0$.}  As $\QQ(X,B)$ has
transcendence degree $n$ over $\QQ(X)$, every $B_i$ is algebraic over
$\QQ(X,f_1,\ldots,f_n)$, i.e., there exist non-zero polynomials
$P_1,\ldots,P_n \in \ZZ[X,W,T]$ such that $P_i(X,f,B_i)=0 \in
\ZZ[X,B]$. View $P_i$ as a polynomial of degree $d_i \in \NN$ in $T$
with coefficients from $\ZZ[X,W]$, and let $Q_i \in \ZZ[X,W]$ be the
(non-zero) coefficient of $T^{d_i}$ in $P_i$. Then $h:=\prod_{i=1}^n
Q_i(X,W)$ is a non-zero polynomial in $\ZZ[X,W]$. By the algebraic
independence of the $f_i$, $h(X,f(X,B))$ is a non-zero polynomial in
$\ZZ[X,B]$; viewing this as a polynomial of degree $e$ in $B$ with
coefficients from $\ZZ[X]$, let $g \in \ZZ[X]$ be any non-zero
coefficient of a monomial $B^\alpha$ of degree $e$.

Now let $\FF$ be any finite field and let $x \in \FF^m$ be such that
$g(x) \neq 0$. Then $q:=h(x,f(x,B))$ is a non-zero polynomial in
$\FF[B]$ of degree $e$. For any $b \in \FF^n$ outside the zero set of
$q$ we have $Q_i(x,f(x,b)) \neq 0$ so that $P_i(x,f(x,b),T) \in
\FF[T]$ has degree $d_i$, for all $i=1,\ldots,n$. Again by
construction, any $b' \in \F^n$ satisfying $f(x,b')=f(x,b)$ satisfies
the system of polynomial equations $P_i(x,f(x,b),b_i')=0$ for
$i=1,\ldots,n$, which has at most $D:=\prod_i d_i$ solutions. We
conclude that the fiber of $f(x,\cdot)$ over $f(x,b)$ has a
cardinality of at most $D$, and therefore
\[ |f(\{x\} \times \F^n)| \geq
\frac{|\{b \in \F^n \mid q(b) \neq 0\}|}{D} \] The
Schwartz-Zippel theorem applied to $q$ shows that the right-hand side
of this inequality is at least $(|\F|^n-e|\F|^{n-1})/D$. From this the
existence of $\gamma$ follows.
\end{proof}

\begin{rem}
The polynomials $P_i,g,$ and $h$ can all be computed effectively,
e.g., using Gr\"obner basis methods \cite{Gathen, LittleCoxOShea}. 
In general, the
running time will depend very strongly on the particular form of the
morphism $f$, but it is independent of the field size $d$, which is
sufficient for our purposes.  It is possible that a more refined
analysis taking into account the structure of $f$ could lead to an
improved performance for certain types of morphisms.
\end{rem}

\begin{rem}
We emphasize that we cannot rule out that the polynomial $g\in\ZZ[X]$
is zero when considered as a polynomial over $\FF$. This can only
happen if all coefficients of $g$ are multiples of the characteristic
$p$ of $\FF$. For this reason, we have to exclude all finite fields
with these characteristics.
\end{rem}

\begin{prop}
Let the $f_i$ be as in Subsection~\ref{ssec:FirstAnalysis}
and $g$ as in Prop.~\ref{prop:FiniteDominant}. 
Assume that the polynomial $g$ is non-zero when
considered over the finite field $\F$. Furthermore, define the set
\[
\Xg := \{x \in \FF^n \mid g(x) \neq 0\}
\]
and for $x \in \Xg$ the set 
\[ 
\Wxg:= \{w \in \FF^n \mid h(x,w) \neq 0 \text{ and }
\eta^x_w \geq 1 \}, 
\]
where $h \in \ZZ[X,W]$ is the polynomial from the proof of
Prop.~\ref{prop:FiniteDominant}. Furthermore, take the constant $D$ as
in the proof.
Then Lemma \ref{Aplemma} can be applied. In particular, the approximative
measurement of Sec.~\ref{Sec app} can be implemented efficiently and
its success probability is bounded from below by a positive and non-zero
constant independent of $d$.
\end{prop}

\begin{proof}
In our application of Prop.~\ref{prop:FiniteDominant} we have $m=n$
and the Jacobian determinant $\det(\partial f_i/\partial B_j)$ is
non-zero as after specializing all $X_i$ to $1$ it is a
non-zero scalar times the Vandermonde determinant
$\det(B_j^{i-1})_{ij}$. This shows that we have a non-zero Jacobian
matrix.  If the image of $g$ in $\FF[X]$ is non-zero then by the
Schwartz-Zippel theorem at least $|\FF|^n-\deg(g)\cdot|\FF|^{n-1}$of
the elements $x \in \FF^m$ lie in $\Xg$, hence we have $| \Xg | \in
O(d^n)$. By the proof of Prop.~\ref{prop:FiniteDominant}, for all $x
\in \Xg$ the set $\Bxg$ from Eq.~(\ref{EEeq2}) contains $O(d^n)$
elements $b \in \F^n$ with $q(b)\not=0$. Since for these $b$ the fiber
of $f(x,\cdot)$ over $f(x,b)$ contains at most $D$ elements, we also
have $O(d^n)$ elements in $\Wxg$. With Rem.~\ref{RemRem} the lower
bound for the success probability follows.

The membership in $\Xg$ can be computed efficiently because we only
have to evaluate $g(x)$. Furthermore, for given $x\in \Xg$ and $w \in
\F^n$ the membership of $w$ in $\Wxg$ can be checked efficiently: By
computing the zeros of the univariate polynomials $P_i(x,w,T)$ in
$\FF$ we find the possible values for each of the $b_i$, and then we
need only to determine\footnote{This can be done more efficiently by
  the replacement of the $P_i$ with a triangular system that can be
  used to find the elements of $\Sxw$ consecutively.}  those
combinations that are mapped to $w$. This also allows us to compute
$\Sxw$ efficiently for $x\in \Xg$ and $w \in \Wxg$.
\end{proof}

Using these results, we show that the success probability of the
approximate measurement is bounded from below by a constant for $n=2$
and fields of characteristic $p=2$. Recall that the first analysis
cannot be applied in these cases since the characteristic is not
strictly greater than the degree.
\begin{exa}
We consider the case $n=2$ and find the two polynomials
\begin{eqnarray*}
P_1(X_1,X_2,W_1,W_2,T)&:=&(-X_1X_2 -X_1^2)T^2 + (2W_2X_1)T +(W_1 X_2 -W_2^2)\\
P_2(X_1,X_2,W_1,W_2,T)&:=&(-X_1X_2 -X_2^2)T^2 + (2W_2X_2)T +(W_1 X_1 -W_2^2)
\end{eqnarray*}
with the leading terms
\begin{eqnarray*}
Q_1(X_1,X_2,W_1,W_2)&:=&-X_1X_2 -X_1^2\\
Q_2(X_1,X_2,W_1,W_2)&:=&-X_1X_2 -X_2^2\,.
\end{eqnarray*}
Therefore, we have
\[
h(X_1,X_2,W_1,W_2)=X_1 X_2(X_1+X_2)^2,
\]
i.e., the polynomial $h \in {\mathbb Z}[X,W]$ is of degree zero 
in $W$ and we have 
\[
g(X_1,X_2)=X_1 X_2(X_1+X_2)^2\,.
\]
Hence, for the maximum degree $n=2$ of the hidden functions we find
polynomials $P_1$ and $P_2$ where $x \in \FF^2$ with $g(x)\not=0$ 
exists for all finite fields $\F$ with $|\F| \geq 3$.
\end{exa}

\section{Conclusion and Outlook}
We have shown that certain instances of the hidden polynomial problem
that are hard on classical computers can be solved efficiently on a
quantum computer for a fixed total degree $n$ and a fixed number $m$
of indeterminates provided that the characteristic of the underlying
field meets certain constraints.

The extension of our results to arbitrary characteristics $p$ of the
field $\F$, to more general algebraic structures, e.g., rings with
Fourier transforms, and the extension to a broader class of functions
such as rational functions are possible objectives of future
research. Additionally, it would be important to find other polynomial
black-boxes with efficient quantum algorithms and to explore if
interesting real-life problems can be reduced efficiently to such
black-box problems.

\subsection*{Acknowledgments}
T.D. was supported by CIFAR, NSERC, QuantumWorks, MITACS and the ARO/NSA
quantum algorithms grant W911NSF-06-1-0379. J.D. was supported
by DIAMANT, a mathematics cluster funded by NWO, the Netherlands
Organisation for Scientific Research.  P.W. gratefully acknowledges
the support by NSF grants CCF-0726771 and CCF-0746600.

\end{document}